\documentclass[mathleft
]{an}
\usepackage{amsmath, amssymb}
\usepackage{graphicx}
\usepackage{times}
\begin{document}

% The following seven commands are intended for editorial usage and should be ignored by
% the author(s).
\Pagespan{1}{}% Document's page range. 
% If second parameter is left empty, the last page is computed automatically.
\Yearpublication{}%
\Yearsubmission{}%
\Month{}%   
\Volume{}%  
\Issue{}% 
% \DOI{This.is/not.aDOI}% 

\title{The Outbursts of Classical and Recurrent Novae}

\author{M.F. Bode}
%Example 
%for footnote, note the usage of the \texttt{fnmsep}
%command as separator between institute number and footnote mark} 

\titlerunning{The Outbursts of Novae}
\authorrunning{M.F. Bode}
\institute{Astrophysics Research Institute, Liverpool John Moores University, Birkenhead, CH41 1LD, UK; mfb@astro.livjm.ac.uk}

\received{}
\accepted{}
\publonline{}


\abstract{In this review, I present our current state of knowledge regarding both Classical Nova and Recurrent Nova systems. Two particular objects (V1974 Cyg and RS Oph) are chosen to illustrate the range of phenomena that may be associated with their outbursts. The wider importance of nova research is emphasised and some of the most pressing unsolved problems are summarised.}

\maketitle

\section{Introduction}

The outbursts of novae have been recorded for over 2000 years (for an historical review, see Duerbeck 2008). It was only in the 1920s during the ``Great Debate''  that it was realised that ``ordinary'' novae such as T Aur (1892 - often seen as the first well-studied nova outburst) were very distinct from ``supernovae'' such as S And (1885 - in M31). Later, Dwarf Novae (DNe) and Classical Novae (CNe) were in turn recognised as rather different beasts and certain of the Classical Novae were also subsequently reclassified as Recurrent Novae (RNe) when a second major outburst was recorded (the earliest example being T Pyx with the 1902 outburst repeating that first noted for this object in 1890).

The watershed in our understanding of the outbursts of CNe (and later RNe) came from a combination of the realisation that cataclysmic variables in general, and CNe in particular, had close binary central systems, together with the suggestion that explosive hydrogen burning was the root cause of the outburst (Kraft 1964). The relationship between DNe, CNe and RNe has long been established. The debate over the relationship of RNe to supernovae continues, and will be touched on here, and in several other contributions in these proceedings.

\section{Classical Novae}

\subsection{Generic properties}

Historically, the primary defining characteristic of the nova outburst is the optical light curve (see Fig.1). Canonically, a rapid rise to maximum is followed by an early decline whose rate defines the ``speed class'' of the nova, with the very fast CNe declining at $> 0.2$ mag d$^{-1}$ (see e.g. Warner 2008). The speed class is in turn related to both the ejection velocity and peak absolute magnitude (in the sense that faster novae have higher ejection velocities and are intrinsically brighter at the peak in their $V$ light output). These properties are of course related to fundamental parameters such as the mass of the accreting white dwarf, and the latter lies behind the Maximum Magnitude-Rate of Decline (MMRD) relations which in the past have been proposed as making CNe potentially important distance indicators. The most widely used MMRD relation is that of Della Valle \& Livio (1995), but this is still subject to significant scatter and uncertainties and is now mainly used as a distance estimator to individual Galactic CNe. The evolution of the CN light curve is also mirrored by a well defined sequence of spectral development comprising in turn the pre-maximum, principal, diffuse-enhanced, Orion, nebular and post-nova spectral stages (see e.g. Duerbeck 2008, Warner 2008).

The central binary comprises a WD primary, of either the CO or ONe type, accreting material from its lower mass companion. The lowest $M_{\rm WD}$ observed in a CN system is $\sim 0.5$M$_\odot$, although selection effects come into play here. The boundary of initial mass between the two WD types is around 1.1 - 1.2 M$_\odot$.  For the observed $P_{\rm {orb}}$ range of  $\sim 1.4 - 8$ hours, the secondary star is a low mass, main sequence object. There are a few longer period systems (e.g. GK Per - see below), but for these, the secondary has to be evolved if it is still to fill its Roche Lobe. The typical accretion rate through the disk is $\dot{M}_{acc}  \sim10^{-9}$ M$_\odot$ yr$^{-1}$ and the luminosity of the quiescent system is $\sim$L$_\odot$ (see e.g. Warner 2008 and references therein).

Around a quarter of CVs contain WDs with a strong enough surface magnetic field to affect the accretion flow markedly. For $B \gtrsim 10^7$G, the field is so strong that the field lines connect to the inter-star stream and no accretion disk forms. The accretion is then predominantly onto one of the magnetic poles of the WD via an accretion column and the source is known as a ``polar'' as a result. For lower field strengths ($10^6 \lesssim B \lesssim 10^7$G) a disk forms, but is disrupted in its inner regions where the WD's magnetosphere controls the flow. Such systems are known as intermediate polars. Among the CNe population, we currently know of one polar (V1500 Cyg) plus around a dozen intermediate polars (including such well studied novae as DQ Her and GK Per - see Warner 2002, 2008). In most polars, as one might expect, the WD's rotation is locked to that of the binary. However, in V1500 Cyg $P_{\rm orb} > P_{\rm rot}$ with the latter increasing such that synchronisation will occur in $\sim 185$ years. It appears therefore that V1500 Cyg has been temporarily de-synchronised by angular momentum transfer during its 1975 outburst. As Warner (2008) points out, the existence of at least 3 other polars with $P_{\rm orb}$ and  $P_{\rm rot}$ differing by $\sim 1\%$ (V1432 Aql, BY Cam and CD Ind) suggests that they might have had nova outbursts in the relatively recent past.

The CN explosion is due to a thermonuclear runaway (TNR) on the WD surface once sufficient H-rich matter has been accreted at a given WD mass for the critical pressure to be attained at the base of the accreted envelope (there are also effects of composition and the rate of accretion itself to be taken into account, and detailed modelling of the TNR relies on accurate determination of nuclear reaction rates - see e.g. Starrfield, Iliadis \& Hix  2008). It is clear nevertheless that TNR models have been very successful in modelling the general form of the luminosity evolution and observed abundances in novae. However, there are still some unsolved details (see below and e.g Jos\'{e} \& Shore 2008). 

The outburst then results in an increase in luminosity typically to $L \sim$few$\times10^4$L$_\odot$ (i.e. $\gtrsim$ L$_{\rm Edd}$ for a 1 M$_\odot$ WD). As fast novae are inherently more luminous, they are more likely to arise from higher mass WDs. Large amounts of mass are ejected at high velocities typically with $10^{-5} \lesssim M_{\rm ej} \lesssim$ few$\times10^{-4}$M$_\odot$ and few$\times10^2 \lesssim v_{\rm ej} \lesssim$ few$\times10^3$ km s$^{-1}$. The heavy element enrichments seen in the ejecta are evidence of mixing between the accreted and the core material. Indeed, CNe are predicted to be the major source of several isotopes in the Galaxy, including $^{15}$N and $^{17}$O (Jos\'{e} \& Hernanz 1998). The inter-outburst period for CN explosions is then thought to be $\sim 10^3 - 10^5$ years, and each system may undergo thousands of outbursts.

As noted below, CNe have been divided into various sub-types, but one often used derives from the observed apparent abundances in the ejecta derived from spectroscopy early in the emission line stage. As noted by Shore (2008) the $CO$ novae show enhanced CNO but generally not heavier elements, and the He/H abundance is near solar. They show a diverse range of light curve behaviour and spectroscopic development and can be prolific dust producers. The $ONe$ or $neon$ novae on the other hand are principally distinguished by the early appearance of the [Ne II] 12.8$\mu$m emission line and often show evidence of depletion of C relative to solar, with associated enhancements of O and Ne. Gehrz (2008) suggests that the most extreme neon novae seem to have much higher ejecta masses than their extreme CO counterparts. As noted by Shore (2008), given the expected rarity of massive progenitors, there has been a surprising number of recent novae of the neon subclass (although we need to be aware of selection effects here of course), including for example V1974 Cyg (see next section). Unlike the CO novae, this subclass appears surprisingly uniform in its characteristics.

The optical light curve is however a misleading indicator of the evolution of the bolometric luminosity of a CN outburst. The existence of a constant bolometric phase of post-outburst development was first proposed following multi-frequency observations of FH Ser (1970 - see e.g. Gallagher \& Starrfield 1976). The initial optical decline is then due to a redistribution of flux to shorter wavelengths due to a decreasing mass-loss rate giving rise to a shrinkage of the effective (pseudo)photosphere at constant bolometric luminosity as steady nuclear burning continues on the WD surface. From simple considerations, Bath \& Harkness (1989) find that the effective photospheric temperature is given by $T_{\rm eff} = {\rm T_0} \times 10^{\Delta V / 2.5}$ K, where $\Delta V$ is the decline in $V$ magnitude from peak and T$_{\rm 0}$ = 8000 K (Evans et al. 2005). This relationship does not take into account contributions from emission lines which may be particularly important at later times, and the maximum effective temperature is reached when the pseudo-photosphere has the same radius as that of the WD. Nevertheless, as Bath \& Harkness note, the explicit dependence of the physical conditions on the decline stage is consistent with the observed relationship of the optical decline to the spectral development noted above.

\begin{figure}
\includegraphics[width=78mm,height=49mm]{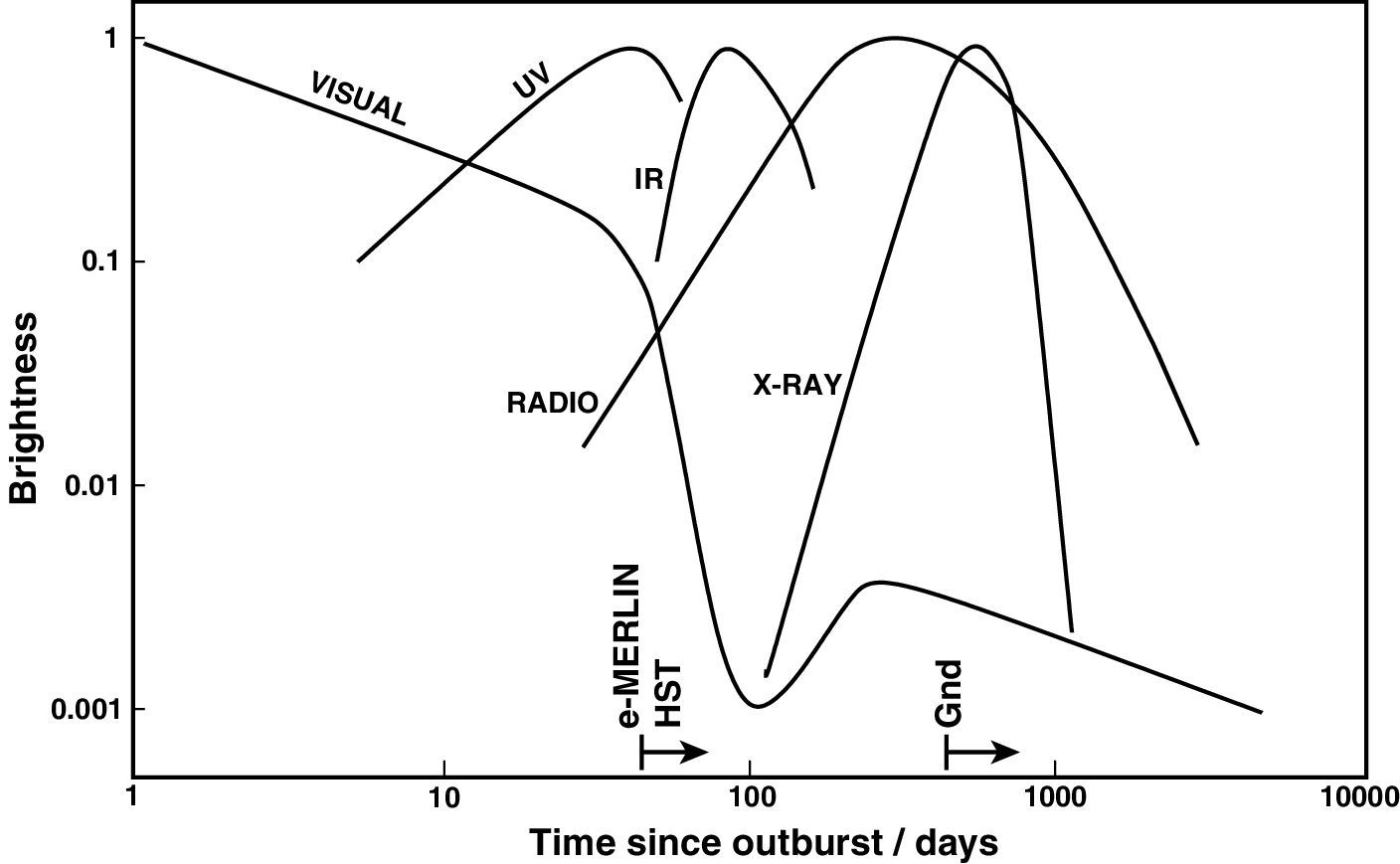}
\caption{Schematic multi-frequency development of a CN outburst with times at which a nebular remnant with expansion velocity of $1000$ km s$^{-1}$ and distance 1 kpc becomes spatially resolved in the radio (e-MERLIN) or optically from space (HST), and on a conventional ground-based optical telescope (Gnd) (from O'Brien \& Bode 2008).}
\label{label1}
\end{figure}

Figure 1 shows the effect schematically for the expected visible and UV behaviour. The subsequent visual light curve break and coincident infrared rise shown here are due to the formation of an optically thick dust shell, covering the whole of the sky as seen from the nova, and which seems to occur preferentially in moderate speed class novae (see e.g. Evans \& Rawlings 2008, Gehrz 2008, who emphasise the importance of novae as real-time laboratories for the investigation of the formation of astrophysical dusts). It may be noted that for $T_{\rm eff} = 5\times10^5$ K, typical of the super-soft source phase, then $\Delta V = 4.5$ mags, and the X-ray development shown schematically here refers to this. However, significant extinction by dust formation prior to this time (typically commencing at $\Delta V \sim 3.5$ in fact) can complicate this picture (further discussion of the SSS source phase in CNe is given below and a full review is given in these proceedings by Hernanz). Finally, the radio light curve arises from thermal (free-free) emission from the expanding ionised gas shell which is initially optically thick and from which the most reliable estimates of ejected mass are derived (Seaquist \& Bode 2008 - see further discussion in Sections 2.2 and 4).

Over 40 CNe now have optically resolved remnants, and 10 have been resolved in the radio (O'Brien \& Bode 2008). Optical spectroscopy of  such resolved remnants has allowed their true geometry to be explored. This in turn has led to detailed modelling of the effects on remnant shaping by the early post-outburst common-envelope phase and has also allowed the refinement of expansion parallax distances with consequent potential improvement to the MMRD (see Bode 2002 and references therein). Observations of extended structure in at least two CNe are however most likely related to their pre-nova evolution.

The very fast neon nova GK Persei rivalled the brightness of Vega at the peak of its outburst in 1901 (see Bode, O'Brien \& Simpson 1994, and references therein). Early observations showed it to possess optical nebulosities on arcminute scales apparently expanding at super-light velocities and subsequently explained as light echoes (Kapteyn 1902). Indeed, it was the first astronomical source in which such motion was observed and one of only three novae where such an effect has been noted (the other two being V732 Sgr (Swope 1940) and V1974 Cyg (Casalegno et al. 2000) - see next section). By 1916 the expanding nebular remnant was observed and it has turned out to be the longest lived and most energetic among the CNe. Indeed, it appears to be more like a supernova remnant (SNR) in miniature but evolving on human timescales (Seaquist et al. 1989). Multifrequency observations and archival research reported in Bode et al. (2004) confirmed the presence of an ancient planetary nebula (PN) pre-dating the 1901 outburst, with an apparent hourglass morphology, ejected by the nova in a previous phase of evolution around $\sim 3\times10^4$ yrs ago (for $d = 470$ pc) and into which the nova ejecta are now running. The 1901 nova outburst was therefore the first of ultimately very many that this system will undergo. Furthermore, from measurements of the proper motion of the central binary we can now understand the asymmetries observed both in the outer (ancient planetary) and inner (1901 nova ejecta) nebulae and their relationship to the longest lived of the light echoes observed over 100 years ago. 

Recently, the IPHAS H$\alpha$ survey revealed the presence of a PN-like nebulosity centred on the position of the subsequent explosion of nova V458 Vul (2007 - Wesson et al. 2008). The nebula has all the hallmarks of a PN that was shed $\sim 14,000$ years ago (for $d = 13$ kpc). The nova flash is now illuminating the nebula and light echo modelling holds out the promise of refining the distance and determining the exact geometry of the PN itself. In the meantime, GK Per and V458 Vul remain potentially important objects, not only for understanding the evolutionary history of CNe, but also for furthering our knowledge of other important astrophysical phenomena (e.g. PN and SNR evolution).

We now know of just under 400 Galactic CNe. Until recent years, the mean detected rate was $\sim$3 per year. Pietsch (these proceedings) now estimates that the observed rate is nearer 8 per year as more systematic and deeper sky survey work is undertaken. However, Liller \& Meyer (1987) estimated that the observable number of CNe with $V > 11$ is around 12 per year, and Warner (2008) suggests that a significant number of bright novae are still overlooked. The discovery rate compares to the total estimated Galactic nova rate of $34^{+15}_{-12}$ yr$^{-1}$ deduced by Darnley et al. (2006) from comparison with the rate of $65^{+16}_{-15}$ yr$^{-1}$ they determined for M31 (where around 800 nova candidates have been catalogued to date -  Pietsch, these proceedings).

In terms of the CN population and its relation to the underlying stellar population, Della Valle et al. (1992) found that fast novae are confined to within $z < 100$ pc of the Galactic plane, whereas slow novae range up to $z \sim 1000$ pc (i.e. the Galactic bulge). They suggested that this may be indicative of two distinct classes of nova progenitor. It is indeed natural to imagine higher mass WDs being more common in the disk than in the bulge and therefore giving rise to a higher incidence of fast nova outbursts as a result. Earlier, Arp (1956) had noted a bimodal distribution of speed classes for novae observed in M31 (although Shafter et al. (2009) have cast some doubt on its reality). Williams (1992) has in turn defined two classes of novae based on their outburst spectra: ``FeII novae'' ($\sim60$\% of the observed total) being slower novae having lower ejection velocities, and ``He/N novae'' being faster novae with higher ejection velocities and encompassing the neon novae.

\subsection{V1974 Cyg - a case study}

By virtue of the rapid response of observers to the notification of its outburst, and the subsequent density of temporal and breadth of wavelength coverage, nova V1974 Cyg still ranks as the best studied CN. Following its discovery on 1992 February 20, it reached $V_{\rm max} \sim 4.2$ mag, was therefore the brightest nova since V1500 Cyg in 1975, and subsequently declined such that $t_3 \sim 35$ days, making it a moderately fast nova. Warner (2008) notes that the amplitude of the outburst was comparatively high, but that the post-nova magnitude was more normal, suggesting a very low state of {\em \.M}$_{\rm acc}$ prior to outburst. The distance to the nova was subsequently determined by a variety of methods to be $\sim 2$ kpc (see below) and the orbital period of the central binary, $P_{\rm orb} =1.95$ hrs (Warner 2008). This is among the shortest known orbital periods for a CN and places it at the lower edge of the $\sim 2-3$ hour period gap for CVs (although there is some debate as to whether there really is a period gap as far as CNe are concerned - see Warner 2002).

Spectroscopy with $IUE$ was conducted from very soon after outburst and caught the nova in the early ``fireball'' stage, first named by Gehrz (1988), and where the effective temperature of the pseudophotosphere drops from $\sim 15,000$ to $< 10,000$K (Hauschildt 2008, Shore 2008). In this stage, captured serendipitously in very few novae to-date, both the lines and continuum are optically thick and the spectrum of the nova resembles that of a Type II supernova with low expansion velocities ($V_{\rm max}\sim4000$ km s$^{-1}$ determined from the UV resonance lines in the case of V1974 Cyg). Emission line profiles showed structure indicating at least two distinct components: relatively featureless high velocity material responsible for the line wings and slower moving knots leading to the multiple peaks seen at smaller displacements from the line centres (Shore 2008). Such structure is obviously formed at very early times in the outburst and the gross features have been explained as being derived during the common envelope phase when the ejecta first overwhelm the secondary (see e.g. Bode 2002 and references therein). Subsequent apparent narrowing of lines in the nebular phase most probably does not reflect a real deceleration of ejected material however, unlike the case of the recurrent nova RS Oph (see below).

Infrared spectroscopy of V1974 Cyg revealed the early emergence of the [Ne II] 12.8$\mu$m emission line, and subsequently those of highly ionised species (``coronal lines''), in particular those of neon, which persisted for a long period thereafter (well over a decade as observed by Spitzer). Indeed, V1974 Cyg is seen as an extreme neon nova (Gehrz 2008 and references therein) with abundance analyses yielding overabundance factors compared to solar of Ne/H $= 71 \pm 17$ for example (Vanlandingham et al. 2005). Considerable velocity structure was also observed in the infrared coronal lines. Saizar \& Ferland (1994) and Hayward et al. (1996) have shown that the structure observed in the [Ne II] and [Ne VI] lines of V1974 Cyg is consistent with an expanding shell comprising dense clumps ($T \simeq 10,000$K) embedded in a much less dense, higher temperature ($T \ge 500,000$K) gas component. 

X-ray observations with the ROSAT PSPC (0.1 - 2.4 keV) began 63 days after outburst and immediately revealed the presence of a relatively hard X-ray source whose observed flux rose between then and day 147 while the spectrum, modelled as emission from a thermal plasma, gradually softened from a derived plasma temperature of $\sim 10$ keV on day 63 to $\sim5$ keV on day 91, being more or less constant at around 1 keV thereafter. Balman et al. (1998) ascribed the origin of the X-rays to shocked emission from $\gtrsim10^{-6}$ M$_\odot$ of gas with shock speeds of 1500-3000 km s$^{-1}$, probably arising from interaction with a pre-outburst, low density wind from the WD. However, another possibility they considered is intra-ejecta shocks as proposed by O'Brien et al. (1994) for the observed early-time hard X-ray emission from the very fast neon nova V838 Her.

The total ejected mass has been determined as $M_{\rm ej} = 3.1\times10^{-4}$ M$_\odot$ for $d =2$ kpc (Hjellming 1996). Such a high mass appears typical of that of the extreme neon novae (Gehrz 2008) and is problematic in terms of the TNR models of this and other objects that predict around an order of magnitude less ejected mass (e.g. Starrfield et al. 1998). It is also interesting to note that V1974 Cyg was also observed in $\gamma$-rays by the COMPTEL instrument on board the Compton Gamma Ray Observatory (CGRO). The failure to detect the $^{22}$Na 1275 keV line in this and other neon novae (Ilyudin et al. 1995) also poses a significant problem for TNR models, particularly in light of the high ejected mass derived observationally (see e.g. Hernanz 2008). It is also puzzling that although the luminosity and ejected mass in V1974 Cyg suggested that the mass density of the ejecta at the deduced condensation radius exceeded the critical density for extensive dust grain condensation, no extensive dust shell formed. As discussed by Gehrz (2008), some additional factor seems to inhibit dust production in extreme neon novae such as this.

The ROSAT observations noted above first showed an indication of the emergence of SSS emission on day 97, ascribed by Balman et al. (1998) to the gradual revealing of continued large-scale nuclear burning on the WD surface as the overlying H column decreased (see also Hernanz, these proceedings). A so-called ``plateau'' phase of the SSS lasted from around 255 days until a decline began somewhere between 511 and 612 days post-outburst. It was also possible to determine the turn-off time from observations of the He II 1640\AA~ line (see Vanlandingham et al. 2003, and Fig. 2). The start of the decline was ascribed to the turn-off of H-burning and on the assumption that the energy radiated by this component thereafter derived from gravitational contraction only, Krautter et al. (1996) then derived an estimated mass for the H-exhausted remnant envelope on the WD surface of $\sim 10^{-5}$ M$_\odot$. Fits to the SED were most satisfactory using a model atmosphere with enhanced O-Ne-Mg abundances (Balman et al. 1998) and the WD mass was estimated to be 1.25 M$_\odot$ by Krautter et al. (1996), which is consistent with expectations for an ONe WD. The hard component was still detectable throughout the SSS phase, but at a very low level. Overall, the ROSAT observations of V1974 Cyg were the most extensive for any nova until those of RS Oph with Swift (see below) and the brightness of the SSS even allowed investigation of dust in the ISM via the observed scattered X-ray halo (Draine \& Tan 2003).

\begin{figure}
\includegraphics[width=78mm,height=54mm]{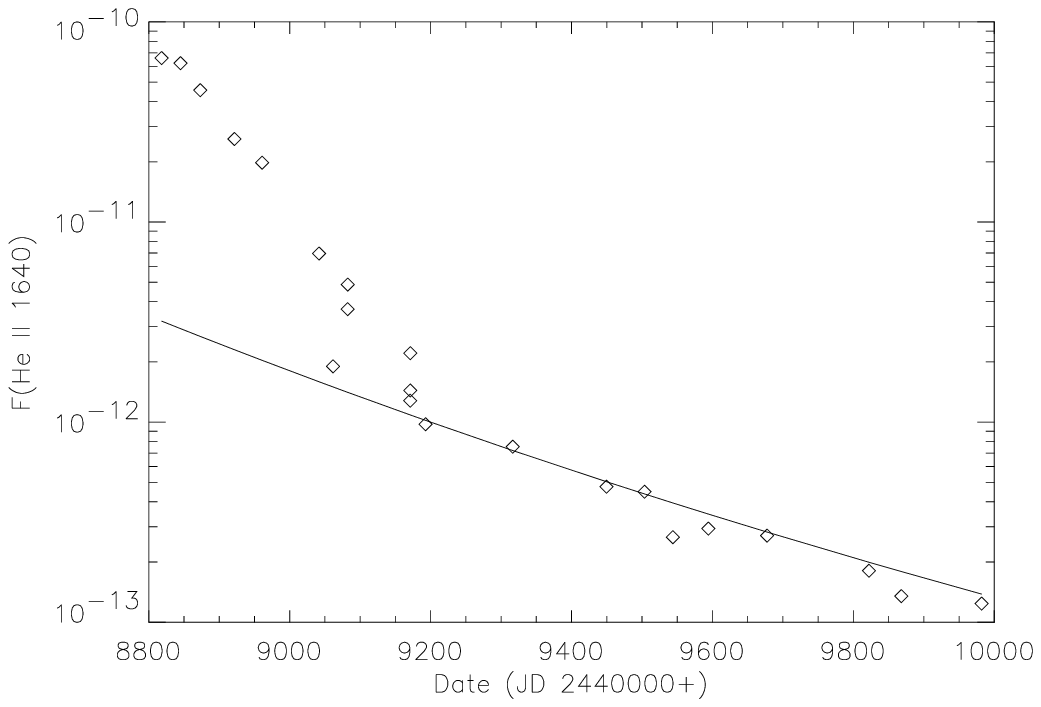}
\caption{He II 1640\AA~ decay during the nebular phase of V1974 Cyg. The transition marks the onset of the SSS turnoff of the central source and the solid line indicates the decline in line flux arising from recombination in uniformly expanding ejecta (from Shore 2008). As Shore notes, with the demise of IUE, suitable surrogates at optical or infrared wavelengths will now have to be used. Possibilities for tracking SSS evolution include [Fe VII] 6089\AA~ and particularly [FeX] 6376\AA~ (see e.g. Ness et al., 2008).}
\label{label2}
\end{figure}

Radio observations were conducted from 21 to 1185 days after outburst and the remnant was first detected with MERLIN after only 80 days (see Seaquist \& Bode 2008, and references therein). These observations allowed fundamental parameters such as ejected mass and distance to be determined, but when compared with mm-submm data it was clear that neither the standard Hubble flow nor variable wind models produced satisfactory results. The radio imagery demonstrated however that there were deviations from spherical symmetry plus evidence for temperature gradients not accounted for in these simple models. 

V1974 Cyg was also the first nova to be observed with HST (Paresce et al. 1995; Krautter et al. 2002 - see Fig. 3) beginning 467 days after outburst. The HST images clearly show a bright elliptical ring-like structure with condensations. Using a combination of HST images and optical and UV spectroscopy, Chochol et al. (1997) concluded that the nebular remnant comprises an outer fast-moving, tenuous, low mass envelope and an inner, low velocity, high mass envelope in the form of an equatorial ring and polar blobs (as seen or inferred in several other CNe). From these observations, they determined an inclination for the ring of 38.7$\pm$2.1 deg (cf. that derived by Krautter et al. of 42 deg), most probably reflecting that of the orbital plane of the central binary, and $d = 1.77 \pm0.11$ kpc. Finally, HST GHRS spectra at 1300 days (Shore et al. 1997) revealed an enhancement of the N/He ratio in a knot and a possible depletion of C with respect to He compared to the general remnant. As the knots are thought to be formed at an early stage of the outburst, such observations of the nebular remnant at late times can potentially elucidate the r\^{o}le of differential mixing of the WD in the accreted layers prior to TNR.

\begin{figure}
\includegraphics[width=82mm,height=31.2mm]{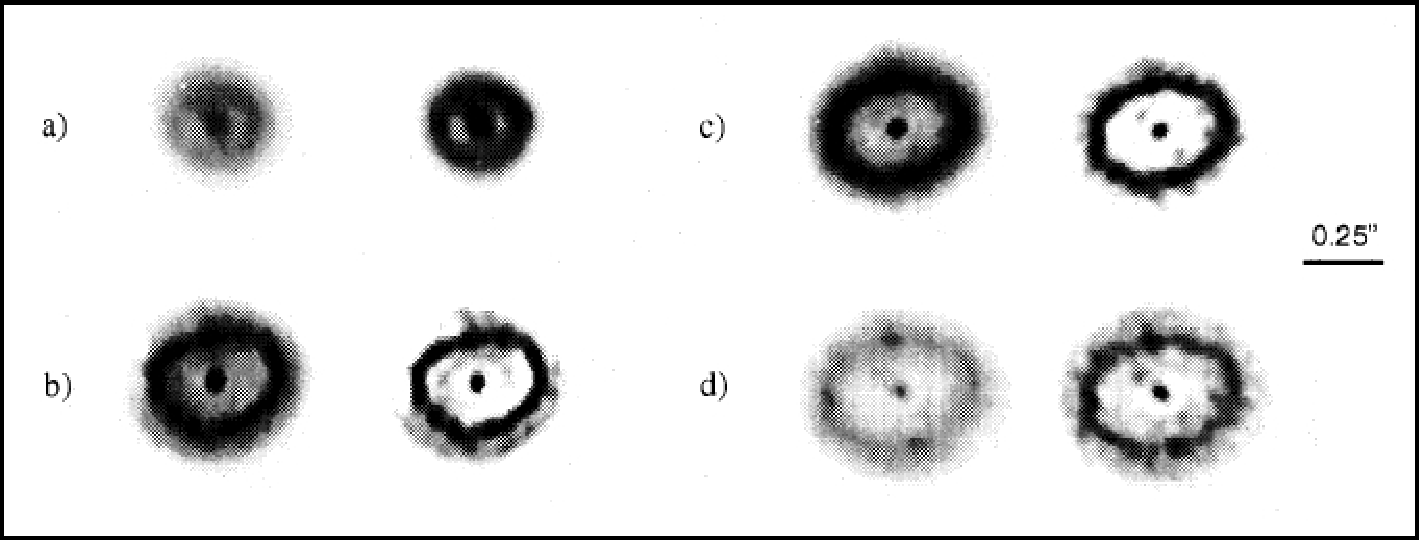}
\caption{The evolution of the resolved shell of nova V1974 Cyg. (a) 278 nm filter (F278M) from May 1993: raw HST images are on the left, deconvolved images on the right: narrow-band [O III] (F501N) images from (b) Jan 1994, (c) Feb 1994, (d) May 1994. From Paresce et al. (1995).}
\label{label3}
\end{figure}

\section{Recurrent Novae}

\subsection{Generic properties}

Unlike CNe, RNe have typical inter-outburst timescales of order decades. 
At present we have secure identifications of a total of only 10 RNe in the Galaxy, although several others are identified in extragalactic systems (see e.g. Pietsch, these proceedings). Anupama (2008) has attempted to classify the Galactic RNe by sub-type, {\em viz.}: \\ The {\em RS Oph/T
  CrB} group  with red giant secondaries, consequent long orbital periods
($\sim$ several hundred days), rapid declines from outburst ($\sim
0.3$ mag day$^{-1}$), high initial ejection velocities ($\ga 4000$ km
s$^{-1}$) and strong evidence for the interaction of the ejecta with
the pre-existing circumstellar wind of the red giant (see next section).\\  The
more heterogeneous {\em U Sco} group with members' central systems
containing an evolved main sequence or sub-giant secondary with an
orbital period much more similar to that in CNe (of order hours to a
day), rapid optical declines (U Sco itself being one of the fastest
declining novae of any type), extremely high ejection velocities
($v_{ej} \sim 10,000$ km s$^{-1}$, from FWZI of emission lines for U
Sco) but no evidence of the extent of shock interactions seen in RS Oph
following outburst (their post-outburst optical spectra resemble the ``He/N''
class of CNe).\\ The {\em T Pyx} group
again comprises short orbital period systems and although their optical spectral
evolution following outburst is similar, with their early-time spectra
resembling the ``Fe II'' CNe, they show a very heterogeneous set of
moderately fast to slow optical light curve declines. This group of systems also seems to show ejected masses similar to
those at the lower end of the ejected mass range for CNe with $M_{\rm
  ej} \sim 10^{-5}$ M$_{\odot}$ (i.e. one to two orders of magnitude
greater than $M_{\rm ej} $ in the other two sub-types of RNe).

RN outbursts are again due to TNR on a WD dwarf surface. Schaefer (2005) has investigated for both T Pyx and U Sco the proposed relationship between inter-outburst timescales and quiescent luminosity that these models predict and has found this to be verified. He then goes on to predict for example that U Sco is due for its next eruption any time now. The short recurrence periods of RNe require high mass WD accretors and
relatively high accretion rates (e.g. Starrfield et al. 1988, Yaron et al. 2005, plus contributions by Hernanz and Kato in these proceedings). Indeed, both
RS Oph and U Sco appear to have WDs near to the Chandrasekhar mass
limit. Unlike CNe where the WD mass is generally thought  to be decreasing (although this is still an interesting open question in fact), the WD mass in both these objects has been proposed as growing
such that they are potential SN Ia progenitors (see e.g. Sokoloski et al. 2006 and Kahabka et al. 1999 respectively, plus next sections). The study of RNe is thus important for several broader fields of
investigation including mass loss from red giants, the evolution of
SNR and the determination of the progenitors of Type Ia SNe.

\subsection{RS Oph - a case study}

RS Oph has had recorded outbursts  in 1898, 1933, 1958,1967, 1985 and 2006, plus probable eruptions in 1907 and 1945. The optical behaviour from one outburst to the next is very similar. The central system comprises a high mass WD in a 455 day orbit with a red giant (M2III) and $d$ (= 1.6$\pm 0.3$ kpc) and $N_{\rm H}$ ($= 2.4\pm0.6 \times 10^{21}$ cm$^{-2}$) are well defined (see Evans et al. 2008, and papers therein).

Optical spectra from outbursts prior to 1985 showed evidence for the existence of a low velocity red giant wind ($u \simeq 20$ km s$^{-1}$) into which the high velocity ejecta ($v_{\rm 0}  \simeq 4000$ km s$^{-1}$) were running. The 1985 outburst was the first to be observed beyond the visual but it was only with the latest eruption on 2006 February 12 that very detailed radio imagery and X-ray observations in particular could be performed. The X-ray evolution was observed from around 3 days after outburst by both Swift (Bode et al. 2006) and RXTE (Sokoloski et al. 2006). These observations revealed a bright, rapidly evolving thermal source consistent with emission from the shocked wind ahead of the ejecta. For Swift, the source was clearly detected in the lowest energy channel of the BAT (14-25 keV) at the time of the outburst itself and source evolution was followed for around 200 days with the XRT (e.g. Page et al. 2008, Bode et al. 2008).

The XRT spectra obtained over approximately the first three weeks post-outburst could be well fitted with a single temperature {\em mekal} model with the interstellar $N_{\rm H}$ fixed and that from the overlying wind allowed to be a free parameter. The shock velocity was derived from the temperature resulting from the fit. The results were consistent with a rapid change at around 6 days after outburst from ``Phase I'' to ``Phase III'' behaviour (Bode et al. 2006) as defined for supernova remnants where the forward shock moves into a $1/r^2$ density distribution (Bode \& Kahn 1985). Subsequently, Nelson et al. (2008), Drake et al. (2009) and Ness et al. (2009) have presented the results of their analyses of grating observations with Chandra and XMM-Newton, which commenced 13.8 days after outburst. These have allowed them to explore in more detail such things as shock evolution and elemental abundances in the emitting gas.

At around 26 days a new, low energy, spectral component rapidly emerged which was initially highly variable and dominated the X-ray emission for approximately the next 70 days (see Osborne, Kato, these proceedings). Indeed, most of the additional count rate resulted from a huge increase in counts below 0.7keV. Here again, Ness et al. (2007) and Nelson et al. (2008) have presented analyses of Chandra and XMM-Newton spectroscopy at several epochs during this SSS phase. The duration of the SSS phase was relatively short, implying that the mass of the WD is high (consistent with the short inter-outburst period - see also Hachisu et al. 2007) and a 35s quasi-periodic modulation was present up to day 65 (Beardmore et al. 2008). The origin of this very short period modulation is still enigmatic. A much longer period oscillation has been observed in V1494 Aql (Drake et al. 2003) and proposed as due to non-radial g$^+$ modes from  the WD. The 35s pulsation seen in RS Oph may however be due to a nuclear burning instability similar to that suggested for (albeit longer period) oscillations in a PN nucleus (Kawaler 1988).

\begin{figure}
\includegraphics[width=78mm,height=54mm]{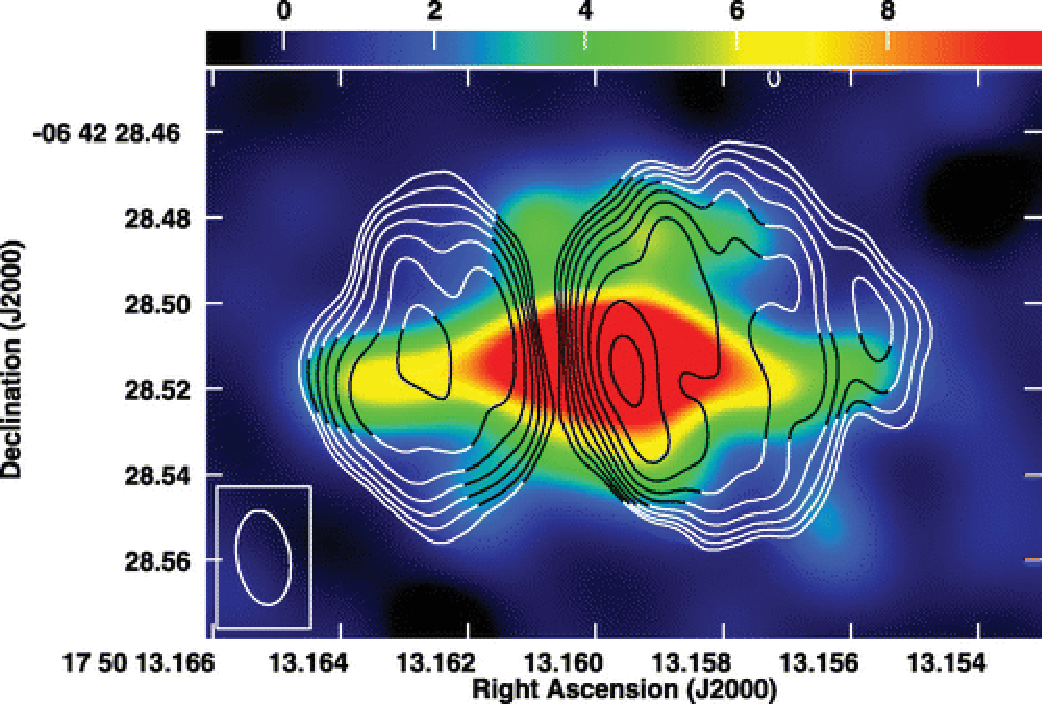}
\caption{Highly collimated outflows and lobes in RS Oph around 50 days from outburst. The contour lines show 1.7 GHz emission (as imaged by the VLBA), which preferentially traces synchrotron-emitting gas. The grey-scale image shows 43 GHz emission (VLA), which preferentially traces thermal plasma. North is up and east is to the left (from Sokoloski et al. 2008).}
\label{label4}
\end{figure}

Turning now to the radio, VLA and MERLIN observations began within a few days of the 2006 outburst. VLBI imaging started on day 13 with the VLBA resolving a partial ring of non-thermally dominated emission of radius around 9 mas, consistent with the dimensions expected for the forward shock at this time as derived from modelling of the X-ray emission (O'Brien et al. 2006, see also Rupen et al. 2008). More extended emission in the form of two lobes gradually appeared to the E then the W and the whole structure eventually bore a close resemblance to that derived from comparatively rudimentary VLBI observations secured by Taylor et al. (1989) on day 77 of the 1985 outburst (Sokoloski et al. 2008; it should also be noted that coincidentally the binary phase was approximately the same at both epochs). Indeed, the suggestion from the 1985 observations that there was a thermal core of emission centred between two non-thermal lobes was borne out by observations of the 2006 outburst (see Fig. 4). Sokoloski et al. (2008) went on to suggest that there were two jets of emission emanating from the central system, directed E-W, and giving rise to non-thermal lobes at the working surfaces of the jets (again, see Fig. 4).

Hubble Space Telescope observations of the expanding nebular remnant were conducted on days 155 and 449 through narrow band filters. Extended structure was particularly evident in [OIII]$\lambda5007$ where two lobes of emission were clearly present (Bode et al. 2007). The spatial extent of the nebula was consistent with that in the radio with expansion velocity $\simeq 3200$ km s$^{-1}$ in the plane of the sky, and no deceleration of the outermost parts apparent between the two epochs. The combination of contemporaneous ground-based optical spectroscopy and modelling of the day 155 image led Ribeiro et al. (2009) to conclude that the lobes arise from a dumbbell-shaped emission region and that there is a region of much higher density material in the centre which in contrast may show some evidence for deceleration. Furthermore, they were able to determine the inclination of the nebula to the plane of the sky as $39^{+1}_{-10}$ degrees and that the West lobe is approaching the observer (in line with conclusions from early-time VLTI infrared interferometry by Chesneau et al. 2007). The true maximum expansion velocities then exceed 5000 km s$^{-1}$.

Bode et al. (2007) and Ribeiro et al. (2009) suggest that the axis of the lobes lies along that of the binary orbit and that there is an enhancement in the density of the pre-existing red giant wind in the equatorial plane of the binary (as has been proposed for such systems by e.g. Mastrodemos \& Morris 1999). Shock propagation is therefore much more rapid in the direction of the orbital pole than in the plane (see also Walder et al. 2008; Drake et al. 2009; Orlando et al. 2009). Whether the collimation of any jet also occurs because of this anisotropic distribution of circumbinary material, or is due to interaction with an accretion disk around the WD, is an open question. The existence of a dense region near the centre of the system which shocks may not entirely traverse would also explain the unexpected survival of significant amounts of dust following the outburst, as revealed by Spitzer (Evans et al. 2007).

\section{Concluding remarks and open questions}

As has been noted above, we have made spectacular advances over the last 50 years in our understanding of the nova phenomenon. We have also realised the significance of their study for diverse branches of modern astrophysics. However, as might be expected, there are many important questions still to answer. The following list is ordered from pre-nova, through outburst, to wider issues. It is not exhaustive, and is inevitably subject to some personal biases, but it hopefully encompasses many of the most pressing problems we currently face:

\begin{itemize}

\item{What is the evolutionary track of a binary to the CN/RN phase? Conversely, can we use novae to help to trace the evolutionary history of binary stars in general? GK Per and V458 Vul may help with the former, and although inevitably rare, there may be more examples to be found. However, studies of novae in extragalactic systems, including the determination of the relationship of nova rates and sub-types to stellar populations, may turn out to be particularly important here.}

\item{Is there a continuum of inter-outburst timescales (and other fundamental properties) from CNe through RNe (at least for the short period sub-types of RNe)? Again, studies of large samples of novae in extragalactic systems such as M31 are likely to help here.}

\item{Do we need to revisit our TNR models particularly in the light of the Swift results? For example, what are the true durations of any super-Eddington and ``plateau'' phases of the bolometric evolution of novae? In this regard, and others, it is a great shame that there is no equivalent of the IUE satellite either currently in operation, or indeed planned for the longer term.}

\item{Related to the previous point, what is the cause of the remarkable variability during the emergence of the SSS in RS Oph, and also, is a nuclear burning instability really the origin of the 35s modulation seen in the XRT data while the SSS was bright?}

\item{There is still a significant (order of magnitude) discrepancy between the ejected mass found from models of the CN outburst compared to that derived from observations (although this seems less of a problem with the U Sco and RS Oph sub-classes of RNe). Consideration of clumpiness in the ejecta helps to a degree, but does not appear sufficient and it may mean we need to re-examine our models.}

\item{Can studies of objects such as RS Oph help to determine more precisely the mechanism of the formation of jets in astrophysical sources? Although examples among the nova population may be rare, and perhaps confined only to the RNe, there is no doubting the wider importance of such work.}

\item{Is there indeed a link between RNe and Type Ia SNe? Although it seems that for systems like RS Oph and U Sco, the mass of the WD is near the Chandrasekhar limit and increasing, there remain some fundamental questions. For example, what is the type of the WD (if ONe, then no SN explosion will occur); can the H in systems with red giant secondaries really be ``hidden'' at the time of any SN outburst; is the population of (appropriate sub-type) RNe sufficient to explain the observed SN Ia rate? A combination of detailed observations of individual Galactic RNe,  coupled with surveys of extragalactic novae, will help to answer these very important points.}

\item{Can the MMRD be refined, possibly by choosing a suitable sub-set of CNe with particular characteristics, and/or by improving the homogeneity and cadence of observations of novae, both Galactic and extragalactic? Although their use as distance indicators for cosmology may have diminished in importance, they may still be very useful to determine the distances to particular extragalactic systems. In addition of course, the MMRD is often used to define the distance to newly discovered novae, and coupled with often poorly determined extinctions, distance estimates  can be very uncertain as a result. The advent of GAIA may prove a watershed here.}

\end{itemize}

%Overall, there is no doubt however that the combination of ever more capable observational facilities, and the application of increasingly sophisticated modelling, will help to ensure that these questions and others are addressed over the coming decades.

%\pagebreak
     
\acknowledgements
I am very grateful to Nye Evans, Sumner Starrfield and an anonymous referee for valuable comments on initial drafts of this manuscript and to Steve Shore and Greg Schwarz for discussions of non-X-ray evidence for the duration of the SSS phase.\\

%\pagebreak

\end{document}